\documentclass[12pt]{revtex4}

\def\be{\begin{equation}}
\def\en{\end{equation}}
\def\bea{\begin{eqnarray}}
\def\ena{\end{eqnarray}}
\def\bec{\begin{equation}\begin{array}{rcl}}

\def\p{\partial}

\def\ve{\varepsilon}
\newcommand{\av}[1]{\langle{#1}\rangle}

\newcommand{\bi}[1]{\mbox{\boldmath$#1$}}

\newcommand{\ten}[1]{\stackrel{\leftrightarrow}{\bi{#1}}}

\usepackage{graphicx}

\begin{document}
\title{Dynamics of 
Binary Mixtures with Ions: \\
Dynamic Structure Factor and 
Mesophase Formation  }  
\author{ Takeaki Araki and Akira Onuki}
\address
{Department of Physics, Kyoto University, Kyoto 606-8502, Japan}

\date{\today}

\begin{abstract}
Dynamic equations are presented 
for polar binary mixtures  containing ions 
 in the presence of the preferential 
solvation. In one-phase states,  we calculate  
the dynamic structure factor 
 of the composition accounting  
for the ion motions.  
Microphase separation can 
take place for sufficiently large solvation 
asymmetry of the cations and  the anions. 
We show  two-dimensional simulation results 
of the mesophase formation with an antagonistic salt, where
 the cations  are hydrophilic  
and the anions are  hydrophobic. 
The  structure factor $S(q)$ 
in the resultant mesophase 
has a sharp peak at an intermediate 
wave number on the order of 
the  Debye-H$\ddot{\rm u}$ckel 
wave number.
As the quench depth is increased, 
 the surface tension 
nearly vanishes in mesophases due to an  
electric double layer. 
\end{abstract}


\maketitle

\section{Introduction}
\setcounter{equation}{0}

Much attention has been paid 
to the phase transition behavior  
arising from  the Coulomb interaction 
among  charged particles 
in various soft matters including  electrolytes,  
 polyelectrolytes, and gels \cite{Levin,Holm,Rubinstein}.   
However, in  most of the theoretical literature, 
the ion-dipole interaction has not been explicitly 
considered, which 
gives rise to a complex structure 
around each ion, called the solvation 
(hydration) shell,  
composed of several   solvent molecules 
(those of the  more polar component 
in  a mixture solvent) \cite{Is}. 
The resultant  solvation 
chemical potential 
$\mu_{\rm sol}^j$ depends 
on the ion species $j$ 
and typically much exceeds the thermal energy $T$. 
It should also  strongly  
depend  on the composition for binary mixtures 
and the polymer volume fraction for 
polymer solutions, so it cannot be neglected at phase transitions 
and around composition  heterogeneities. 
Recently, including the preferential solvation effect,  
several theoretical 
groups  have begun to investigate  
the ion effects  in electrolytes 
\cite{OnukiJCP04,OnukiPRE,Tsori,Roij,Andelman1}, 
polyelctrolytes\cite{Onuki-Okamoto}, 
and ionic surfactants \cite{OnukiEPL}.

We mention some  experiments directly related 
to our theory. 
First, many authors have long observed 
salt-induced phase separation or homogeneization 
in aqueous binary mixtures, where the phase behavior 
is strongly altered even by  a  small amount of an  
 salt   \cite{polar}.  
Second, we   mention  a number of 
 observations 
of salt-induced aggregates 
in near-critical binary mixtures \cite{So,Wagner,third}, 
where the cations and anions 
are both hydrophilic. 
In one-phase states  \cite{So,Wagner},  heterogeneities 
 extending  over a few micrometers 
have been   detected       by light scattering 
with addition of a  salt 
(for example, $17~$mass$\%$ NaBr 
in  mixtures of  
H$_2$O+3-methylpyridine (3MP) \cite{Wagner}).  
In two-phase state \cite{third}, a macroscopic thin plate 
 has been observed at  a liquid-liquid interface, 
which  presumably consists of aggregates 
of  solvated ions.  
Third, we mention  
 recent small-angle neutron scattering 
experiments  by Sadakane {\it et al.} \cite{Sadakane,S1}. 
They  added sodium tetraphenylborate 
NaBPh$_{4}$   at $ 100$ mM to a     mixture of D$_2$O 
and 3MP to  find  a  peak   
at an intermediate wave number $q_m$($\sim 0.1~$\AA$^{-1}$). 
The peak height of the SANS  intensity was much 
enhanced with formation of periodic structures.  
Their salt is composed of hydrophilic  Na$^+$ 
and hydrophobic  BPh$_{4}^-$.  Furthermore,  
the mixture exhibited colors 
changing dramatically on approaching 
the criticality at 
low salt contents $(\sim 10~$mM).

Hydrophilic  and hydrophobic ions 
interact    differently   with the 
composition   fluctuations 
in mixtures of water+less polar component. 
They behave antagonistically in the presence of 
the composition fluctuations. We may predict 
formation of a large electric 
double layer  at liquid-liquid interfaces 
much reducing the surface tension  
and formation of  mesophases   
for sufficiently large solvation asymmetry 
\cite{OnukiJCP04,OnukiPRE}. 
However, we  do not know the details of 
the phase transition of binary mixtures 
with an antagonistic salt. 
 Originally,  
Nabutovskii {\it et al.}\cite{Russia} 
 pointed out a possibility of 
mesophases  in electrolytes 
assming a coupling between the composition and 
the  charge density in the free energy.

In Section 2,   we will present   a short summary 
of the statics of  
binary mixtures containing ions 
accounting for  the preferential solvation.   
In Section 3,  dynamic equations for such systems 
will be given and, 
as an application,  the dynamic scattering amplitude 
will  be calculated.  In Section 4, 
we  will then  numerically examine 
the mesophase formation 
induced by  antagonistic ion pairs.

\setcounter{equation}{0}
\section{Ginzburg-Landau free energy}
\subsection{Electrostatic and solvation interactions}

We consider a polar binary 
 mixture  containing a small amount of  salt. 
The composition 
of a  water-like  component  is  written as $\phi$. 
The cation and anion  densities are 
written as  $n_1$ and $n_2$ with charges 
$Z_1e$ and $Z_2e$. In the monovalent case 
we have $Z_1=1$ and $Z_2=-1$.  They are sufficiently dilute and 
 their  volume fraction is negligible.  
The chrage density is given by 
$e(Z_1n_1+Z_2n_2)$.   
The variables  $\phi$, $n_1$, and $n_2$ vary smoothly 
in space. The Boltzmann constant will be set equal to unity. 
As the geometry 
of our  system, our fluid  is 
between  parallel metallic 
plates in the region $0<z<L$. 
The lateral dimensions in the $xy$ plane 
are much larger than $L$.  
The surface charges on the plates can 
give rise to an applied electric field. 
In the following theory, 
we fix the charges on  the plates 
such that  their  electrostatic energy 
is kept fixed \cite{NATO}.

The Ginzburg-Landau free energy functional 
of our system is written as 
$F=\int d{\bi r}f$ with  the free energy density 
\cite{OnukiJCP04,OnukiPRE},  
\begin{eqnarray}
&&{f} = f_0(\phi,T) +
\frac{TC}{2}|\nabla\phi|^2+ \frac{\varepsilon{\bi E}^2}{8\pi }
\nonumber\\
&& + T\sum_j
\bigg [\ln (n_jv_0) -1-  g_j \phi\bigg]n_j.  
\end{eqnarray} 
The first two terms constitute  the usual 
Ginzburg-Landau free energy density. 
The chemical part 
$f_0=f_0(\phi,T)$ depends on $\phi$ and $T$ 
and the coefficient  $C$ of the gradient term is of the order 
$a^{2-d}$ in $d$ dimensions, where 
$a$ is  the molecular radius. 
The third  term is the 
electrostatic free energy, where 
${\bi E}=-\nabla\Phi$ is the electric field and 
the electrostatic potential $\Phi$ satisfies  
 the Poisson equation 
\be 
-\nabla\cdot\ve(\phi)\nabla \Phi=
 4\pi e(Z_1n_1+Z_2n_2).  
\en 
The dielectric constant  $\ve(\phi)$ can   depend  on 
the composition $\phi$. In our previous work 
the linear composition dependence  
\be 
\ve(\phi)=\ve_0 + \ve_1 \phi
\en 
has been assumed, where $\ve_0$ is the 
dielectric constant of the less polar component 
and $\ve_0+\ve_1$ is that of the 
water-like component. 
In such cases $\ve_0>0$ and $\ve_1>0$.  
This linear form approximately holds 
in some polar binary mixtures \cite{Debye}.
The last term in (2.1) consists of the entropic part 
and  the solvation contribution  
of the ions, where the parameters $g_j$  represent  
 the solvation strength. The choice of the volume 
$v_0$ is arbitrary and is taken 
to be the solvent molecular volume (see (5.1) below). 
In this work we neglect  the image interaction  
arising  from inhomogeneous dielectric constant 
or from nonvanishing $\ve_1$ in our 
theory \cite{OnukiJCP04,OnukiPRE,Levin1}. 
The interfacial  ion distribution is then 
produced by the  preferential solvation 
among the ions and the mixture solvent. The image 
interaction is weakened with increasing the salt density 
and/or approaching the critical point.

With (2.1) we may  calculate 
the chemical potentials  
$h=\delta F/\delta \phi$ 
and $\mu_j=\delta F/\delta n_j$. 
They  are written as  
\bea 
\frac{h}{T} 
&=& \frac{f_0'}{T} - C\nabla^2\phi  -\frac{\ve_1{\bi E}^2}{8\pi T}
- \sum_j g_jn_j,\\
\frac{\mu_j}{T}
&=& \ln (n_jv_0) - g_j\phi +\frac{1}{T}Z_je  \Phi,
\ena 
where $f_0'= \p f_0(\phi)/\p \phi$ and  
$\ve_1=\p \ve/\p \phi$.  
If the system is in equilibrium, 
$h$, $\mu_1$, and $\mu_2$ 
are homogeneous constants. 
When the system undergoes a macroscopic 
phase separation with a planar interface 
separating polar and less polar regions,  
we may   calculate the interface 
profiles of the composition and the  ions \cite{OnukiPRE}. 
In equilibrium  the composition difference $\Delta\phi$ and 
the potential difference $\Delta\Phi$ satisfy 
\be 
e\Delta\Phi= T(g_1-g_2)\Delta\phi/(Z_1+|Z_2|),
\en    
from the charge neutrality in the bulk 
regions. The  
 $\Delta\Phi$ is called the Galvani potential difference 
in electrochemistry \cite{Hung,Osakai}.

The solvation free energy may be written as 
 $f_{\rm sol}= \sum_j \mu_{\rm sol}^j(\phi)n_j$,  
where  $\mu_{\rm sol}^j(\phi)$ is the  solvation 
chemical potential of the ion species 
$j$.  It is   assumed to depend on $\phi$ as  
\be 
\mu_{\rm sol}^j(\phi) =\mu_{\rm sol}^j(0)   -Tg_j\phi.
\en  
Here the first term in the right hand side 
 gives a  contribution linear in $n_j$ in $f_{\rm sol}$ 
and is not written in $f$ in (2.1), while the second term 
 yields the solvation  coupling terms in $f$ 
between the ions and the composition. 
We remark on the magnitude of $g_j$. 
In aqueous mixtures,   
it is  positive for hydrophilic  ions and  
negative for hydrophobic ions. 
In two-phase coexistence,  
the difference of the solvation 
chemical potential between the two phases  is 
given by  
$\Delta\mu_{\rm sol}^j= 
Tg_j\Delta\phi,$ 
which is identical to the standard 
Gibbs transfer free energy (per particle) in 
electrochemistry \cite{Hung,Osakai}.  Data of  
$\Delta\mu_{\rm sol}^j$ are available  
for water-nitrobenzene at room temperatures 
in strong segregation 
(where $\Delta\phi\cong 1$).
 For example,  $\Delta\mu_{\rm sol}^j/T=g_j\Delta\phi =$  
13.6 for Na$^+$, 15.3 for Li$^+$,  26.9  for Ca$^{2+}$, 
11.3 for Br$^-$, and 7.46 for I$^-$ as examples of 
hydrophilic ions, while it is 
$-14.4$ for 
BPh$_4^-$ (tetraphenylborate)  
as an example of hydrophobic ions. The anion BPh$_{4}^-$    
consists  of  four phenyl rings bonded to an ionized 
boron,    acquiring   strong hydrophobicity.  
Note that  Sadakane {\it et al.} \cite{Sadakane,S1} 
used NaBPh$_{4}$.  
Thus   the preferential 
 solvation effect can  be  very strong. 
However,  it  
has mostly been neglected in 
theories of electrolytes 
and  soft matters, though it  strongly influences  
 phase transitions in such systems.

When phase-separation occurs macroscopically, 
a liquid-liquid interface appears. 
If the space dependence is along the $z$ axis, 
the surface tension 
 is expressed as \cite{Onuki-Okamoto} 
\be 
\gamma=  2\int dz(f_g - f_{el}),
\en 
where $f_g= C|\nabla\phi|^2/2$ is the gradient 
free energy density 
and $f_{el}= \ve{\bi E}^2/8\pi$ is the 
electrostatic free energy density. 
Up  to linear order in the ion densities, 
we  may also derive the expression, 
\be 
\gamma\cong \gamma_0-T\Gamma+ \gamma_{el},
\en  
where $\gamma_0$ is the surface tension without ions, 
$\Gamma$ is the surface adsorption of 
ions, and $\gamma_{el}$ is the electrostatic contribution 
given by $\gamma_{el}= -\int dz f_{el}<0$. 
For antagonistic salts with large $|g_i|$, 
$|\gamma_{el}|$  is much amplified  
due to the electric double layer at the interface.

\subsection{Structure factor in one-phase states}

In our previous papers  \cite{OnukiJCP04,OnukiPRE}, 
we examined 
the structure factor $S(q)= \av{|\phi_{\small{\bi q}}|^2}$ 
 of the composition 
fluctuations with wave number $q=|{\bi q}|$ 
in one-phase states with salt, where $\phi_{\small{\bi q}}$ 
is the Fourier component of 
the composition deviation 
$\delta\phi({\bi r})= \phi({\bi r})-\av{\phi}$ 
with wave vector $\bi q$. 
Hereafter  $\av{\cdots}$ 
denotes the thermal average. 
We readily obtain  $S(q)$  
if  the  fluctuation contributions to   $F$ are calculated 
in the  bilinear order. The resultant 
free energy part is written as $\delta F$. 
The thermal fluctuations obey the Gaussian distribution 
$\propto e^{-\delta F/T}$ in equilibrium in the mean-field theory. 
Hereafter we consider the monovalent case $Z_1=-Z_2=1$, where 
the average ion densities 
are written as $\av{n_1}=\av{n_2}=n_e$.

From (2.1) some calculations give   
\bea
{\delta F} & =&\frac{T}{2}  \sum_{\bi q}  \bigg [
({\bar{r} + Cq^2})  |\phi_{\small{\bi q}}|^2+
\frac{4\pi \ell_B }{q^2}|\rho_{\small{\bi q}}|^2\bigg]\nonumber\\
&&+  T  \sum_{\bi q} 
\sum_{j} \bigg [
\frac{1}{2{n_e}} |n_{j{\small{\bi q}}}|^2 -
g_j n_{j {\small{\bi q}}}\phi_{\small{\bi q}}^*\bigg ] ,
\ena
where   
 $n_{j{\small{\bi q}}}$ and  
$\rho_{\small{\bi q}}$ are  the Fourier components 
of  $n_j({\bi r})$   
and  $\rho({\bi r})\equiv n_1({\bi r})-n_2({\bi r})$, 
respectively, and $\ell_B= e^2/\ve T$ is the 
Bjerrum length.  We define 
\be 
{\bar r}= \p^2 f_0(\phi)/\p\phi^2.
\en 
The average composition 
$\av{\phi}$ is simply written as $\phi$. 
Here we may 
treat $\ve$ as a constant when  we 
 treat  the small thermal fluctuations. 
By minimizing $\delta F$ 
with respect to $n_{j{\small{\bi q}}}$ at fixed $\phi_{{\small{\bi q}}}$, 
we obtain $\delta F/T=\sum_{\bi q} |\phi_{\small{\bi q}}|^2/2S(q)$
with 
\be
\frac{1}{S(q)}
={\bar r}- (g_1+g_2)^2\frac{n_e}{2}  + C q^2 
\bigg[1-  \frac{\gamma_{\rm p}^2\kappa^2}{\kappa^{2}+q^2}\bigg] , 
\en
where 
$\kappa= (8\pi  \ell_Bn_e)^{1/2}$ is the 
Debye wave number and  the parameter 
\be
\gamma_{\rm p}= (16 \pi C \ell_{B})^{-1/2} 
|g_1-g_2|
\en
represents asymmetry of the solvation 
of the two ion species. 
The structure factor thus obtained is 
analogous to that for weakly charged 
polyelectrolytes \cite{Onuki-Okamoto,Lu1,Lu2}.

The second term 
in the right hand side of (2.12)  gives rise to 
a shift of the spinodal curve \cite{polar}. 
For example, if the cations and anions are 
hydrophilic and $g_1\sim g_2\sim 15$, the shift term 
is of order $-500 n_e$ and its magnitude can be 
appreciable even for $v_0n_e\ll 1$.  
On the other hand, $\gamma_{\rm p}$ can be increased 
for antagonistic salts composed of  hydrophilic 
and hydrophobic ions \cite{OnukiJCP04,OnukiPRE,Sadakane,S1}.  
From the last term  in (2.12) a Lifshitz point appears at   
$\gamma_{\rm p}=1$. For $\gamma_{\rm p}>1$,  
$S(q)$ exhibits a peak 
at an intermediate wave number 
 $q_{\rm m}$. Since  the derivative 
of the right hand side of 
(2.12) with respect to $q^2$ vanishes at $q=q_m$, 
we find   
\be 
q_{\rm m}= ( \gamma_{\rm p}-1)^{1/2}\kappa  .
\en 
The peak height is given by $S(q_{\rm m})=1/({\bar r}-r_{\rm m})$, where 
\be 
r_{\rm m} = (g_1+g_2)^2\frac{n_e}{2}  
+ C(\gamma_{\rm p}-1)^2\kappa^2. 
\en 
For  ${\bar r}<r_{\rm m}$, 
mesophase formation takes place, 
as will be studied in Section 4.

\section{Dynamics}
\subsection{Dynamic equations for composition, 
ions, and velocity}
\setcounter{equation}{0}

We present the dynamic equations 
for $\phi$, $n_1$, $n_2$, and the velocity field $\bi v$ \cite{Onukibook}.
The fluid is assumed to be incompressible and isothermal. 
That is, we require 
\be 
\nabla\cdot{\bi v}=0
\en 
and treat 
the mass density $\rho_0$  
and the temperature $T$ as constants. 
Then  $\phi$ and $n_j$ obey  
\begin{eqnarray}
&&\frac{\partial\phi}{\partial t}+
\nabla\cdot(\phi\mbox{\boldmath $v$})= 
L_0 \nabla^2\frac{h}{T},\\
&&\frac{\partial n_j}{\partial t}+
\nabla\cdot(n_j\mbox{\boldmath $v$})= D_j
\nabla\cdot n_j\nabla \frac{\mu_j}{T} \nonumber\\
&&\hspace{-1mm}
= D_j\nabla\cdot\bigg[{\nabla{n_j}}
-\frac{Z_je}{T}  n_j{\bi E} -g_jn_j\nabla\phi\bigg],
\end{eqnarray}
where $h$ and and   $\mu_j$ are given  in (2.4)  and  (2.5),  
$L_0$ 
is the kinetic coefficient 
(with $L_0/v_0$ representing 
a diffusion constant), 
   and   $D_1$ and $D_2$ are 
 the ion diffusion constants in the solvent. 
The momentum equation is expressed as 
\be 
\rho_0  \frac{\partial{\bi v}}{\partial t}=
-\nabla p_1 - \nabla\cdot 
 {\ten \Pi} + \eta_0\nabla^2{\bi v} ,
\en
The first term on the right hand side   ensures 
the incompressibility condition (3.1) 
and $p_1$  satisfies  
\be 
\nabla^2p_1= -\sum_{\alpha\beta}
\nabla_\alpha\nabla_\beta \Pi_{\alpha\beta}, 
\en 
where  $\nabla_\alpha=\p/\p x_\alpha$ 
with $x_\alpha=x,  y, z$.   We introduce 
the reversible stress tensor 
${\ten \Pi}=\{\Pi_{\alpha\beta}\}$ ($\alpha,\beta=x,y,z$) in the form, 
\be 
\Pi_{\alpha\beta}=
{T}C  \nabla_\alpha\phi \nabla_\beta\phi
- \frac{\ve}{4\pi}E_\alpha E_\beta. 
\en 
where the first term is well-known 
in  critical dynamics \cite{Onukibook} 
and the second term is a part of  
the Maxwell stress tensor  
(with its diagonal part being included in  
$p_1$) \cite{Landau}.

We determine $\ten \Pi$ from the relation, 
\be 
\nabla \cdot {\ten{\Pi}} = 
 \phi \nabla h +\sum_j n_j\nabla \mu_j. 
\en 
If  the above relation  holds, 
the total free energy 
$F_{\rm T}=\int d{\bi r}[f+ \rho_0{\bi v}^2/2]$ 
including the fluid kinetic energy 
changes in time as 
\be 
\frac{d}{dt} F_{\rm T}= -\int d{\bi r} [\dot{\epsilon}_\phi 
+\dot{\epsilon}_{\rm vis}+ 
\dot{\epsilon}_{\rm ion}],
\en 
where  the terms in the brackets  are 
the heat production rates in the bulk given by 
\bea  
\dot{\epsilon}_\phi=L_0|\nabla h|^2, \quad 
\dot{\epsilon}_{\rm vis}= 
\eta_0\sum_{\alpha\beta}|\nabla_\alpha v_\beta|^2, 
\nonumber\\ 
\dot{\epsilon}_{\rm ion}=\sum_j {D_j}n_j |\nabla \mu_j|^2/T. 
\ena
The surface terms are omitted in (3.8). 
Owing to $dF_{\rm T}/dt\le 0$, the system tends to equilibrium 
if there is no externally applied flow. 

In our dynamic equations we neglect the random source 
terms \cite{Onukibook}, 
which are related to the 
transport coefficients $L_0$, $D_j$, 
and $\eta_0$ via the fluctuation-dissipation 
relations. They are needed to 
describe the dynamics of the 
thermal fluctuations and to 
calculate the time correlation functions.

\subsection{Stokes approximation}

Without macroscopic flow, the viscous motion  
 of ${\bi v}$ is much faster 
than  the diffusive motions of 
 $\phi$ and $n_j$. 
Here  $L_0/v_0$, $D_1$, and 
$D_2$ are estimated by the Stokes formula 
$(D_j \sim T/6\pi\eta_0 a_j$ with $a_j$ being 
the molecular size), so they are  
much smaller than 
the kinematic viscosity $\eta_0/\rho_0$. 
Then we may well neglect the acceleration 
of the velocity in (3.4)  to obtain \cite{Onukibook} 
\be 
v_\alpha({\bi r})= 
\int d{\bi r}' \sum_\beta  {\cal T}_{\alpha\beta}
({\bi r}-{\bi r}')X_\beta({\bi r}')
\en 
where $X_\alpha({\bi r})= 
-\sum_\beta 
\nabla_\beta \Pi_{\alpha\beta}(\bi r)$ 
is the force density acting 
on the fluid 
and  $ {\cal T}_{\alpha\beta}
({\bi r})$ is the Oseen tensor. 
This Stokes  approximation has been used 
in numerical analysis of spinodal 
decomposition in the literature \cite{Koga}.  
The  free energy 
$F=\int d{\bi r}f$ 
changes in time as 
$dF/dt= -\int d{\bi r} [\dot{\epsilon}_\phi 
+\dot{\epsilon}_{\rm vis}+ 
\dot{\epsilon}_{\rm ion}]\le 0$ as in (3.8), 
where  $\dot{\epsilon}_{\rm vis}$ is replaced by 
\be 
\dot{\epsilon}_{\rm vis}
=\sum_{\alpha}X_\alpha  
v_\alpha 
\en
Here  $\int d{\bi r}\dot{\epsilon}_{\rm vis}\ge 0
$ from the expression (3.10).

\subsection{Ionic local equilibrium}

The composition evolution  
can be much slower than the ionic motions 
particularly near the critical point. 
In such cases, the ion distributions are 
expressed in terms of  $\phi$ and $\Phi$  as  
\be 
n_j=  n_j^0 \exp(g_j \phi -Z_j e\Phi/T),
\en
where the coefficient 
 $n_j^0$ is determined 
from the conservation of the ions 
$\int d{\bi r}n_j({\bi r},t)=$const. 
In numerical analysis 
this approximation is 
convenient to examine the 
mesophase formation for large $g_j$.

\section{Relaxation of the thermal composition 
fluctuations}
\setcounter{equation}{0}

\subsection{Time-correlation function }

In this section, 
we  calculate the time-correlation 
function of the Fourier components 
of the composition fluctuations, 
\be 
{ G}(q,t)
= \av{ \phi_{\small{\bi q}}(t) 
\phi_{\small{\bi q}}(0)^*},
\en 
in one phase states. 
This  function can be measured 
by dynamic scattering. 
It is of interest 
how it  relaxes on approaching 
the spinodal point and how it 
is influenced by the ion diffusion. 
The thermal hydrodynamic fluctuations  
are governed by the linearized hydrodynamic equations of (3.2) 
and (3.3)  with random 
source terms added.  That is, they obey 
linear  Langevin equations \cite{Onukibook}. 
In this section,  without 
explicit introduction of  the noise terms, 
we will calculate 
the time-correlation functions 
of the form  $\av{ {\cal A}_{\small{\bi q}}(t) 
\phi_{\small{\bi q}}(0)^*}$ with $t>0$, where  
${\cal A}= \phi, n_1$, and $n_2$. 
 We also assume that the cations and the 
anions  have  the same diffusion constant 
or  $D_2=D_1$, which much simplifies 
the calculation.

From (3.3) ${G}(q,t)$ 
obeys the linear  equation, 
\be 
\bigg
[\frac{\partial }{\partial t} +
\Gamma_0(q)\bigg] 
{G}
=  L_0q^2 \bigg[
g_1  {G}_1+ g_2{G}_2 \bigg],
\en 
where  ${G}(q,t)$ is written as $ {G}$ 
and 
$\Gamma_0(q)$ is the decay rate without ions,    
\be 
\Gamma_0(q)= 
L_0 q^2 ({\bar r}+Cq^2).
\en 
Here we write 
${G}_1\equiv \av{ n_{1{\small{\bi q}}}(t) 
\phi_{\small{\bi q}}(0)^*}$ and  
${G}_2\equiv \av{ n_{2{\small{\bi q}}}(t) 
\phi_{\small{\bi q}}(0)^*}$, where  
$n_{1{\small{\bi q}}}$ 
and $n_{2{\small{\bi q}}}$ are  the Fourier 
components of $n_1$ and $n_2$.  
The equations for the combinations 
${G}_1\pm {G}_2$   read 
\bea
\bigg
[\frac{\partial }{\partial t} +
D_1q^2\bigg] ( {G}_1 + {G}_2)
=  n_e (g_1 + g_2) D_1 q^2  {G},
\\
\bigg
[\frac{\partial }{\partial t} +
D_1(q^2+\kappa^2) \bigg]( {G}_1 -{G}_2)
= n_e  (g_1 - g_2)D_1 q^2   {G}.
\end{eqnarray}
Use has been made of the fact that 
the Fourier component of the electric potential 
is $\Phi_{\bi q}= 4\pi e(
n_{1{\bi q}}-n_{2{\bi q}})/\ve q^2$ 
from the Poisson equation (2.2), where  
the dielectric constant $\ve$ 
may be treated as a constant. 
The convective terms in (3.2) and (3,3) 
vanish in the linear order without 
velocity gradient.

It is convenient to  calculate 
the  Laplace transformation 
\be 
{\hat G}(q,\Omega)= \int_0^\infty dt e^{-\Omega t} {G}(q,t).
\en  
The Fourier transformation 
$I(q,\omega)= \int_{-\infty}^{\infty}  dt 
e^{-i\omega t}G(q,t)$ is related to 
${\hat G}(q,\Omega)$ by 
\be 
I(q,\omega)= 2{\rm Re}[{\hat G}(q,i\omega)],
\en 
where 
${\rm Re}[\cdots]$ denotes taking the real part. 
Some calculations give the following expression,  
\be 
{\hat G}(q,\Omega)= S(q)\bigg[ 
\Omega + \frac{L_0q^2/S(q)}{1+Z(q,\Omega)
} \bigg]^{-1}.
\en 
The ionic correction 
$Z(q,\Omega) (\propto n_e)$ 
depends on $q$ and $\Omega$ as 
\bea 
{Z}(q,\Omega) &=& \frac{1}{2}n_eL_0q^2 
\bigg [ \frac{(g_1+g_2)^2}{\Omega+D_1q^2}\nonumber\\
&&\hspace{-1.5cm}+ \frac{(g_1-g_2)^2q^2}
{[\Omega+D_1(q^2+\kappa^2)](q^2+\kappa^2)}\bigg]. 
\ena 
In deriving (4.8) and (4.9) 
use has also been made of 
the static relations,
\bea 
&&\av{(n_{1{\bi q}}+n_{2{\bi q}})\phi_{\bi q}^*}
= n_e(g_1+g_2)S(q), \nonumber\\
&&(1+\kappa^2/q^2)\av{(n_{1{\bi q}}-n_{2{\bi q}})\phi_{\bi q}^*}
= n_e(g_1-g_2)S(q),
\ena 
 which follow from (2.10).
These equal-time correlation functions 
appear in the Laplace transformation of $S(q,t)$ 
in the presence of  the random source terms. 
The presence of $Z(q,\Omega)$ in (4.8) 
makes the relaxation of $G(q,t)$ complicated.

\subsection{Relaxation near the spinodal point}  

We obtain 
the exponential relaxation,
\be 
G(q,t)\cong S(q)e^{-\Gamma(q) t},
\en  
near the spinodal point. 
Here the decay rate $\Gamma(q)$ is assumed to 
be much smaller than $D_1q^2$.   
Then  we may set $\Omega=0$ in $Z(q,\Omega)$ to find  
\be 
\Gamma(q)\cong \frac{L_0q^2/S(q)}{1+ n_e B(q)}, 
\en 
where $B(q)= Z(q,0)/n_e$ is written as 
\be 
B(q)= \frac{L_0}{D_1} 
\bigg [\frac{(g_1+g_2)^2}{2}
+ \frac{(g_1-g_2)^2q^4}{2(q^2+\kappa^2)^2} \bigg] .
\en 
If $|g_1|$ and $|g_2|$ 
are very large,  the ionic correction  $n_e B(q)$ 
can be noticeable   
even for $v_0n_e\ll 1$.  
For $\gamma_{\rm p}>1$, 
$\Gamma(q)$ tends to zero first 
at $q=q_m$ on approaching the spinodal point.

\subsection{Long wavelength limit} 

In dynamic light scattering experiments, 
we should consider the long wavelength limit, where  we set 
 $q \ll \kappa$, $\Gamma_0(q) \cong D_\phi q^2$, and  
$Z(q,\Omega) \cong \alpha D_1 q^2/(\Omega+D_1q^2)$ with 
\bea 
&&D_\phi= L_0{\bar r},\\
&& \alpha= L_0 (g_1+g_2)^2n_e/2D_1.
\ena
Here  $D_\phi=\lim_{q\to 0}\Gamma(q)/q^2$ is the diffusion constant 
of the composition in the long wavelength limit 
without ions. The dimensionless parameter 
$\alpha$ is proportional to $n_e$ 
and increases  steeply with increasing $n_e$ 
for $g_1+g_2 \gg 1$. 
In dynamic light scattering without ions, 
 $D_\phi$ tends to zero near the critical point 
(being given by the Kawasaki formula $T/6\pi \eta_0\xi$ 
with $\xi$ being the correlation length)\cite{Onukibook}.  
In this limit we obtain  
\bea 
\frac{{\hat G}(q,\Omega)}{S(q)}
&=&\frac{\Omega+ (1+\alpha)D_1q^2}{\Omega^2+ 
{\cal D} q^2\Omega + D_1D_\phi q^4} \nonumber\\
&=& 
\frac{\beta}{\Omega +D_- q^2}+\frac{1-\beta}{\Omega +D_+q^2},
\ena 
where ${\cal D}=(1+\alpha)D_1+D_\phi$ in the first line. 
The two new diffusion constants 
$D_-$ and $D_+$ in the second line 
are expressed as 
\be 
D_{\pm}= \frac{{\cal D}}{2} \pm \frac{1}{2} 
\sqrt{{\cal D}^2-4D_\phi D_1} .
\en  
The partition coefficient $\beta$ is 
 of the form 
\be 
\beta 
=\frac{1}{2}+ \frac{(1+\alpha) D_1-D_\phi}{2(D_+-D_-)}.
\en 
The inverse Laplace transformation 
of the second line of (4.15) yields 
the time-correlation function exhibiting 
 a double-exponential decay, 
\be 
\frac{{G}(q,t)}{S(q)}=  \beta e^{-D_-q^2t}+
(1-\beta)  e^{-D_+q^2t}. 
\en 
(i) For very small ion concentrations 
there can be  the situation where $\alpha\ll 1$ and  
$D_1\alpha\ll D_\phi < D_1$. In this case 
we have $D_-\cong D_\phi$ and $D_+ \cong D_1$ 
with $\beta= 1- \alpha D_\phi/(D_1-D_\phi)^2+
\cdots$. 
(ii) We may suppose the case 
$D_s \ll D_1\alpha$ . 
In this case we have 
$
D_- \cong D_\phi/(1+\alpha),\quad 
D_+\cong D_1(1+\alpha),
$ 
with  $\beta=1-[(2\alpha-1)/2(1+\alpha)^2] 
D_\phi/D_1+ 
\cdots$.

\section{Simulations at the critical composition}
\setcounter{equation}{0}

We numerically examine 
phase ordering with a strongly  antagonistic salt 
at the critical composition $\av{\phi}=1/2$.  
The  spatial dimensionality $d$ has been 
equal to three so far. However, 
we here present  preliminary simulation results 
in  two dimensions.

\subsection{Numerical method}

In our simulation,  
we choose 
the chemical free energy density 
$f_0$  in  (2.1) in 
the Bragg-Williams form,
\be
\frac{v_0}{T}f_0 =  \phi\ln\phi
+(1-\phi)\ln(1-\phi)+\chi\phi(1-\phi),   
\en 
where  $v_0= a^d$ is the solvent molecular 
volume and   
 $\chi$ is the interaction parameter dependent on $T$.  
The parameter ${\bar r}$ in (2.9) is given by 
${\bar r} =[1/{\phi(1-\phi)}-2\chi]/v_0$. 
Space and time will be 
measured in units of $a$ and 
\be 
t_0= v_0 a^2/L_0,
\en 
where $L_0$ is the kinetic coefficient in (3.2). 
Integration of the dynamic equations is performed 
on a  $256\times 256$ square lattice, 
so the system is in the region $0<x,y<256a$.
Supposing the monovalent case,  we set 
\bea 
&&v_0 C={a^2}, \hspace{2mm}
g_1=-g_2=15, \nonumber
\\
&&
\ve_1=0, \hspace{2mm}
\ell_{\rm B}=3a,\hspace{2mm}
  \frac{\eta_0}{T}= \frac{0.16a^4 }{L_0}.  
\ena 
Then we obtain  $\gamma_{\rm p}\cong 2.44$ from (2.11) and  
 mesophases are realized with increasing $\chi$. 
These values of $g_1$ and $g_2$ are realistic 
in view of the data of the Gibbs transfer free energy,  
as discussed below (2.7). The correlation length is 
 defined by  $\xi= [Cv_0/|4-2\chi|]^{1/2}$, 
which is equal to $a$ for  $\chi=2.5$.

 The velocity field $\bi v$ is determined  
by the Stokes approximation 
(3.10) and the ion densities $n_1$ and $n_2$ 
by the Poisson-Boltzmann expressions (3.12) 
(the latter being justified 
 in the limit $D_j\rightarrow \infty$). 
In the dynamic equation (3.2) for 
$\phi$  we put a  random source term to calculate 
the structure factor, 
\be
\frac{\partial\phi}{\partial t}+
\nabla\cdot(\phi\mbox{\boldmath $v$})= 
L_0 \nabla^2\frac{h}{T} -\nabla\cdot {\bi j}_R,
\en 
Here ${\bi j}_R$ is the random diffusion flux characterized by 
\be 
\av{j_{R\alpha}({\bi r},t)j_{R\beta}({\bi r}',t')}
=2{\tilde L}\delta_{\alpha\beta}
\delta({\bi r}-{\bi r}')\delta(t-t').
\en   
where $\alpha, \beta= x, y$. 
The noise strength  $\tilde L$ should be equal to 
the kinetic coefficient $L_0$ 
to ensure the equilibrium 
distribution ($\propto e^{-F/T}$). 
In this paper, however, 
we set ${\tilde L}=10^{-8}L_0$ to 
detect   the  composition patterns  
unambiguously. In one phase states, 
$\phi$ remains nonvanishing 
due to ${\bi j}_R$, yielding a structure factor 
 proportional to the mean field structure factor 
 $S(q)$ in (2.12), where   the proportionality 
constant  is ${\tilde L}/L_0=10^{-8}$ 
(not shown here). In two phase states, ${\bi j}_R$ 
serves to trigger phase ordering, yielding 
a structure factor composed of 
the domain contribution.
The same structure factor 
follows even if we set ${\bi j}_R={\bi 0}$ 
in the course of domain growth. 
It is worth noting that  
 the random source terms 
are mostly  neglected in the literature of 
phase ordering dynamics \cite{Onukibook}.

In our simulations we start with  the initial condition 
$\phi({\bi r},0)=1/2$ at $t=0$. 
Small disturbances of $\phi$ are subsequently 
produced by  the small 
random flux ${\bi j}_R$ in (5.4), 
which grow into patterns in two-phase states. 
For $g_1=-g_2$ and $\av{\phi}=0.5$, 
use of (2.15) yields  the linear instability criterion,  
\be 
2-\chi< \frac{1}{2}Ca^2 (\gamma_{\rm p}-1)^2\kappa^2,
\en 
where  the right hand side is 
$78v_0n_e$ from (5.3). 
 Hereafter the Debye wave number is 
$\kappa= 8.7 n_e^{1/2}$ with $n_e=\av{n_1}=
\av{n_2}$ being  the average ion density.

\subsection{Mesophase formation in shallow quenching}

Here we study the phase ordering 
at the solvent criticality 
$\chi=2$ and $\av{\phi}=1/2$, 
where instability occurs for $n_e>0$. 
In Fig.1, we show the time evolution of a  
normalized characteristic domain size 
$2\pi/a q_{\rm p}(t)$ for various $n_e$. 
In terms of the time-dependent 
structure factor $S(q,t)= 
\av{|\phi_{\bi q}(t)|^2}$ we define   
\be 
q_{\rm p}(t)=\sum_{\small{\bi q}}q
S(q,t)
/\sum_{\small{\bi q}} S(q,t).  
\en 
In Fig.1,  $q_{\rm p}(t)$ tends to a constant 
expressed as $9.22 n_e^{1/2}$ at long times.  
It nearly coincides with 
$q_m= (\gamma_{\rm p}-1)^{1/2}\kappa
= 10.4 n_e^{1/2}$ in (2.14). 
In Fig.2, the steady-state structure factor $S(q)$ 
is given for three ion densities, where 
all the   curves  arise  from the domain structure and are   
not affected by the  small noise term in (5.4).   
Our $S(q)$ exhibits  a sharp peak  at $q = q_m$ 
and a second peak  at $q = 3q_m$. 
The peak height at  $q = q_m$ 
 is a constant of order unity nearly independent of $n_e$. 
This can be explained as follows. 
It is known that  a domain structure 
gives the structure factor of the form 
$S(q)\cong  (\Delta\phi)^2\ell^{d}
S^*(q\ell)$, where $\Delta\phi$ 
is the composition difference between the two phases, 
$\ell$ is the domain size, and $S^*(x)$ is a scaling function. 
In our two-dimensional case, we have $\Delta\phi \propto n_e^{1/2}$ 
and $\ell \sim 2\pi/q_m \propto  n_e^{-1/2}$, 
so $S(q) \sim S^*(q\ell)$.

In Fig.3, we display $\phi(\bi{r},t)$ 
 and $n_1(\bi{r},t)$ at $t=3000t_0$  
for  $v_0 n_e =0.0005$ (left)  and 0.0025 (right).  
In Fig.4,  we present cross sections 
of $\phi$, $n_1$, $n_2$ in the upper panel and 
those of the gradient free energy  
 $f_g=TC|\nabla\phi|^2/2$ and the 
electrostatic energy 
$f_e=\ve{\bi E}^2/8\pi$, 
and their difference  in the lower panel. 
These quantities vary 
mildly without sharp interfaces 
as  functions of $x$ at fixed $y=64a$.  
We notice  that the difference $f_g-f_e$  
is small. In Fig.5, their space averages, 
$\av{f_g}=\int d{\bi r}f_g/V$ and 
$\av{f_e}=\int d{\bi r}f_e/V$,  
are demonstrated to be nearly 
the same at long times, where 
$V$ is the system volume. 

%1
\begin{figure}
\includegraphics[width=0.6\textwidth,bb=0 0 686 479]{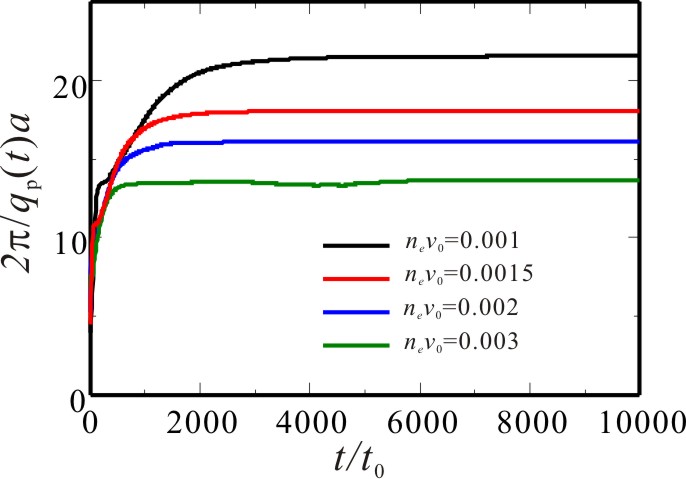}
\caption{
Characteristic domain size $2\pi/q_m(t)a$ vs time 
at the solvent criticality 
for $v_0 n_e= 0.001, 0.0015,0.002$, and $0.003$.
The saturated value of $q_{\rm p}(t)$ nearly coincides
 with $q_m$ in (2.14). 
}
\label{fig1}
\end{figure}
%2
\begin{figure}
\includegraphics[width=0.6\textwidth, bb=0 0 650 459]{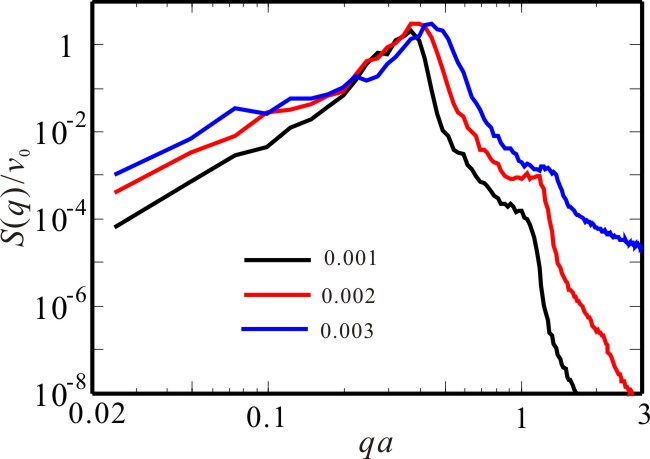}
\caption{Steady-state structure factor  $S(q)$  of 
the composition for $v_0 n_e= 0.001,0.002$, and $0.003$, 
where the the solvent is at  
the criticality ($\chi=2$ and $\av{\phi}=1/2$). 
}
\label{fig2}
\end{figure}
%3
\begin{figure}
\includegraphics[width=0.9\textwidth, bb= 0 0 1024 784]{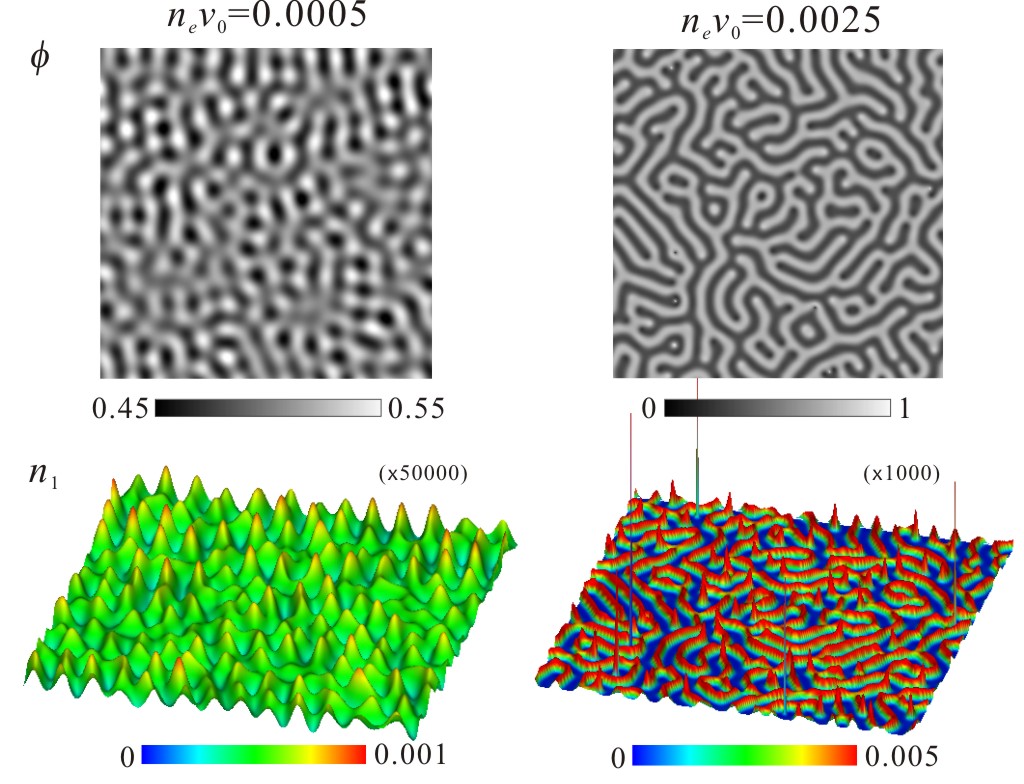}
\caption{
Patterns of $\phi(\bi{r},t)$ (top) 
and $n_1(\bi{r},t)$ (bottom) at $t=3000t_0$ 
for  $n_ev_0=0.0005$ (left) and 0.0025 (right) 
at the solvent criticality. 
}
\label{fig1}
\end{figure}
%4
\begin{figure}
\includegraphics[width=0.6\textwidth, bb= 0 0 654 962]{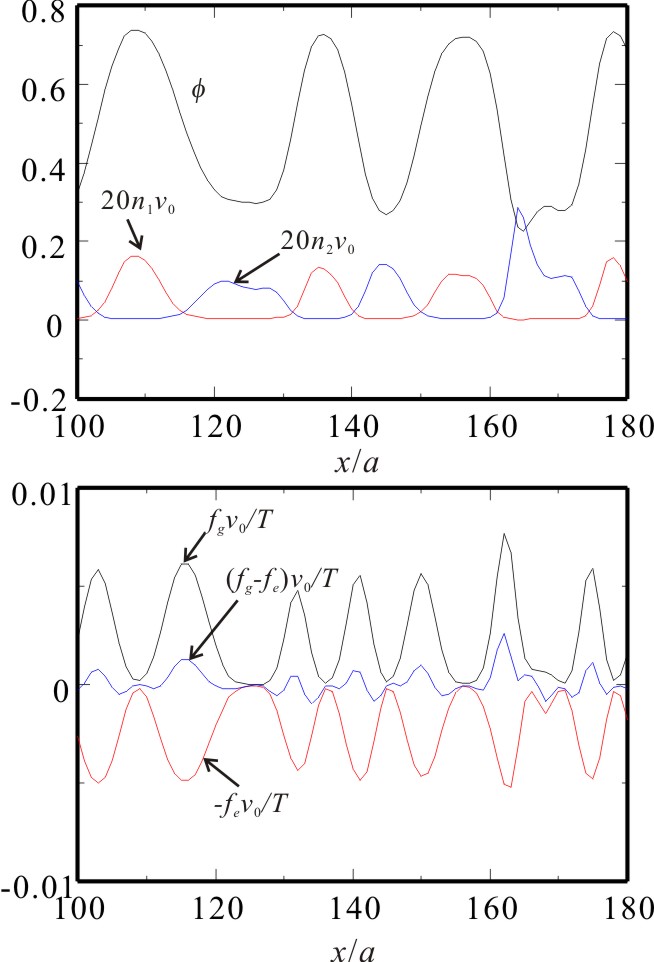}
\caption{
 Cross sections 
 of $\phi$, $v_0 n_1$ and $v_0 n_2$ (top) 
and those of  
$f_gv_0/T$, $-f_e v_0/T$, and  $(f_g-f_e) 
v_0/T$ (bottom) for  $n_ev_0=0.0025$ 
in the region $100<x<180$ at $y=64$, 
where the solvent is at the criticality.  
Use is made of the  data producing 
 the right images in Fig.3.   
}
\end{figure}
%5
\begin{figure}
\includegraphics[width=0.6\textwidth,bb= 0 0 680 481]{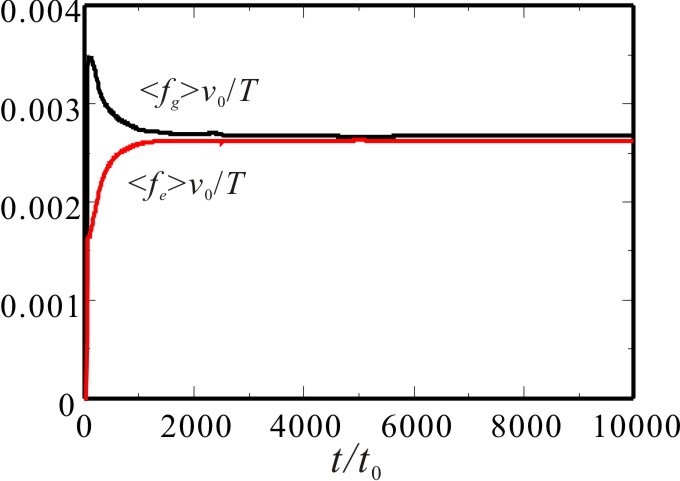}
\caption{Space averages of  
 $f_gv_0/T =v_0C|\nabla\phi|^2/2$ and 
$f_ev_0/T =v_0\ve|\nabla\Phi|^2/8\pi T$ 
vs time $t$ for  $n_ev_0= 0.0025$, 
where the solvent is at the criticality. 
}
\end{figure}

We argue why 
$\av{f_g}\cong \av{f_e}$ holds in steady states 
in weak segregation. 
If the  ion density is small 
at shallow quenching, 
the composition is weakly segregated 
and is composed of the Fourier components 
with $q=|{\bi q}|\cong q_m$. 
As in the weak segregation 
case of block copolymers \cite{Ohta}, 
the deviation 
$\delta\phi= \phi-\av{\phi}$ is expressed as 
\be
\delta\phi=\sum_{\small{\bi q}}A_{\bi q}
e^{i{{\bi q}\cdot{\bi r}}}, 
\en
where the coefficients 
$A_{\bi q}$ are sharply peaked 
at  $q=q_m$. With this form,  
the space average of $f_g$ is written as 
\bea 
\frac{\av{f_g}}{T}
&=&\frac{C}{2V} \sum_{\small{\bi q}}q^2
|A_{\bi q}|^2\nonumber\\
&\cong& \frac{1}{2}C 
{q_m^2} \av{\delta\phi^2},
\ena 
where $ \av{\delta\phi^2}=\sum_{\small{\bi q}}
|A_{\bi q}|^2/V$.  
Linearlizing (2.2) and (3.12) 
with respect to $\delta\phi$ in the monovalent case, 
we obtain the electric potential 
\cite{Onuki-Okamoto},  
\be 
\Phi=\frac{T}{2e}\sum_{\small{\bi q}} 
\frac{(g_1-g_2)\kappa^2}{q^2+\kappa^2}
A_{\bi q}e^{i{{\bi q}\cdot{\bi r}}}, 
\en
From  (2.13) the average electrostatic energy 
is written as   
\bea
\frac{\av{f_e}}{T}
&=&\frac{1}{2V} \sum_{\small{\bi q}} 
\frac{C\gamma_{\rm p}^2\kappa^4q^2}{(q^2+\kappa^2)^2}|A_{\bi q}|^2 
\nonumber\\
&\cong& 
\frac{C\gamma_{\rm p}^2\kappa^4q_m^2
}{2(q_m^2+\kappa^2)^2}\av{\delta\phi^2}  .
\ena
From $q_{m}=({\gamma_{\rm p}-1})^{1/2}\kappa$ in (2.14), 
we find  $\av{f_g}\cong 
\av{f_e}$.

\subsection{Mesophase formation in  deep quenching}

%6
\begin{figure}
\includegraphics[width=0.6\textwidth,bb = 0 0 657 477]{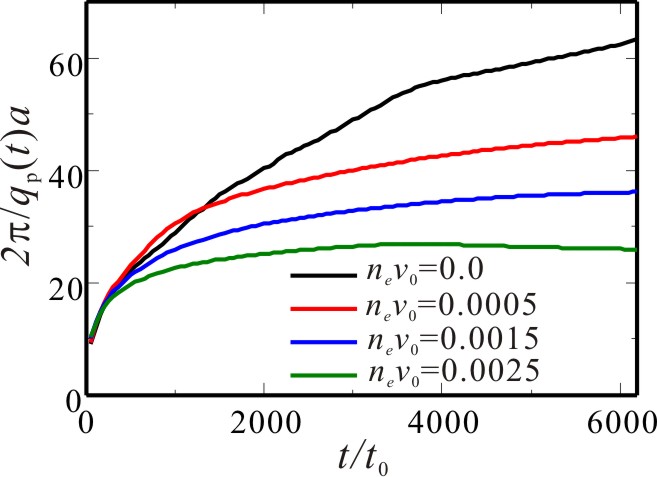}
\caption{
Characteristic domain size $2\pi/q_m(t)a$ vs time 
in deep quenching 
with  $\chi=2.5$ and $\av{\phi}=1/2$ for 
 $v_0 n_e=0,  0.0005, 0.0015$, and $0.0025$.}  
\end{figure}
%7
\begin{figure}
\includegraphics[width=0.6\textwidth,bb= 0 0 682 481]{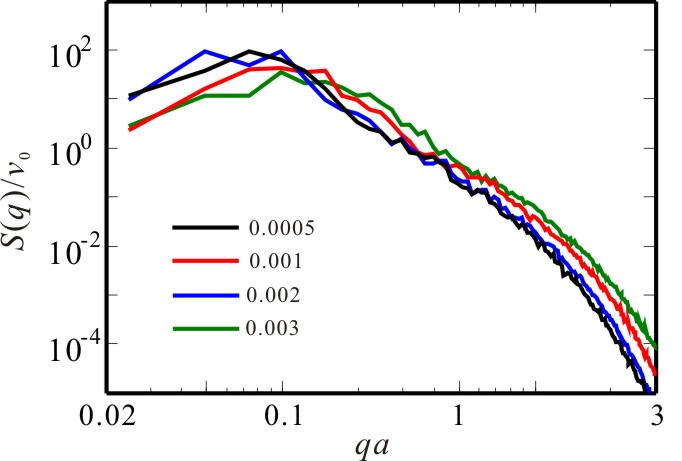}
\caption{
Structure factor  $S(q)$ in deep quenching 
with  $\chi=2.5$ and $\av{\phi}=1/2$ for 
 $v_0 n_e=0,  0.0005, 0.001, 
0.0015$, and $0.0025$.
The numbers in the figure 
denote  $n_ev_0$. 
}
\end{figure}

Next we examine the case 
of deep   quenching by setting $\chi=2.5$ with 
$\av{\phi}=0.5$, where the interface thickness 
is $\xi=a$. In Fig.6, 
we show  the time evolution 
of  the characteristic domain size 
$2\pi/q_{\rm p}(t)$, where $q_{\rm p}(t)$ 
is defined by (5.7). 
For $v_0n_e=0.0025$ the domain size 
approaches a constant, while 
for  $v_0n_e=0.0015$ and $0.0005$ 
 its growth still  continues  in 
the simulation but is extremely slow 
at the end of the simulation 
$(t/t_0=6000$).  In Fig.7, 
the structure factor $S(q)$ is shown
 for $v_0 n_e=0, 0.0005, 0.0015$, and 0.0025. 
The structure factor 
around the peak is of order 
$100 v_0$ and is much larger 
than the thermal level.

In Fig.8, we display $\phi(\bi{r},t)$ 
 and $n_1(\bi{r},t)$ at $t=3000t_0$  
for $v_0 n_e =0.0005$ (left)  and 0.0025 (right).  
As a marked feature for $v_0 n_e =0.0005$,  the 
cations (anions)  are confined in 
the water-rich (water-poor) regions. 
Because of the small ion density here, 
the ions  change discontinuously at the interfaces 
and are homogeneously distributed 
in the  preferred domains. 
On the other hand, for $v_0 n_e =0.0025$, 
the ions are localized near the interfaces. 
In Fig.9, we show cross sections 
of $\phi$, $n_1$, $n_2$ (top) 
and those of $f_g$, $-f_e$, 
 and $f_g-f_e$ (bottom). 
We can see electric double layers 
at the interfaces in accord with 
the theory \cite{OnukiPRE}. 
The difference $f_g-f_e$  turns out to be 
 small in steady states. 
In Fig.10, we demonstrate that 
their space averages    nearly coincide 
at long times. From (2.8) we recognize that the surface 
tension $\gamma$  nearly vanishes  
in steady states.

%8
\begin{figure}
\includegraphics[width=0.9\textwidth,bb= 0 0 948 1009]{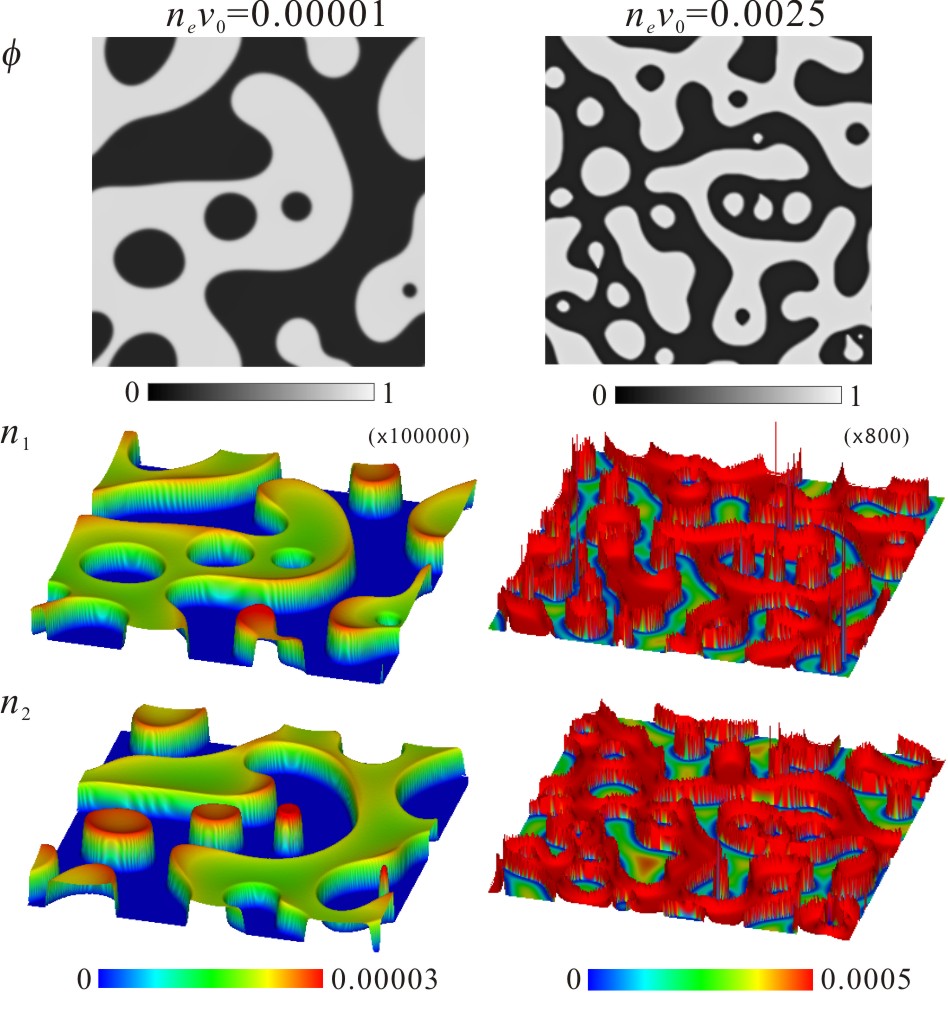}
\caption{
Patterns of $\phi(\bi{r})$ (top),  
$n_1(\bi{r})$ (middle)  
and $n_2(\bi{r})$ (bottom) in deep quenching 
at $t=3000t_0$ for 
 $n_ev_0=0.00001$ (left) and 0.0025 (right), 
where  $\chi=2.5$ and $\av{\phi}=1/2$. 
}
\end{figure}
%9
\begin{figure}
\includegraphics[width=0.6\textwidth,bb= 0 0 661 962]{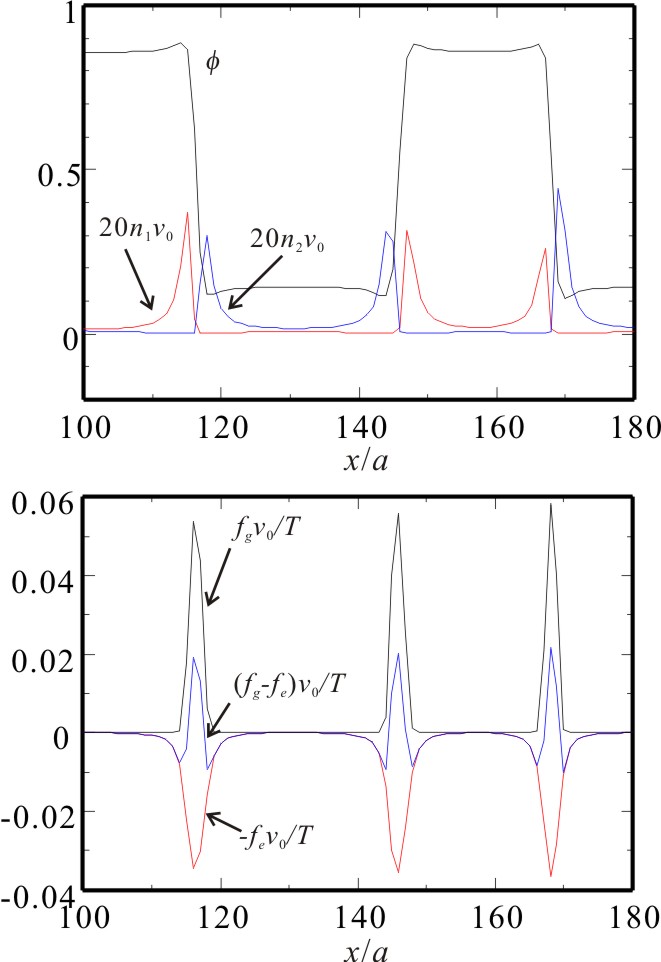}
\caption{
Cross sections 
 of $\phi$, $v_0 n_1$ and $v_0 n_2$ (upper plate) 
and those of $f_gv_0/T$, $-f_e v_0/T$, and  $(f_g-f_e) 
v_0/T$ (lower plate) for   $n_ev_0=0.0025$ 
in the region $100<x<180$ at $y=64$, 
where  $\chi=2.5$ and $\av{\phi}=1/2$. 
Use is made of the data producing  
 the right images in Fig.8.  
}
\end{figure}
%10
\begin{figure}
\includegraphics[width=0.6\textwidth,bb= 0 0 619 480]{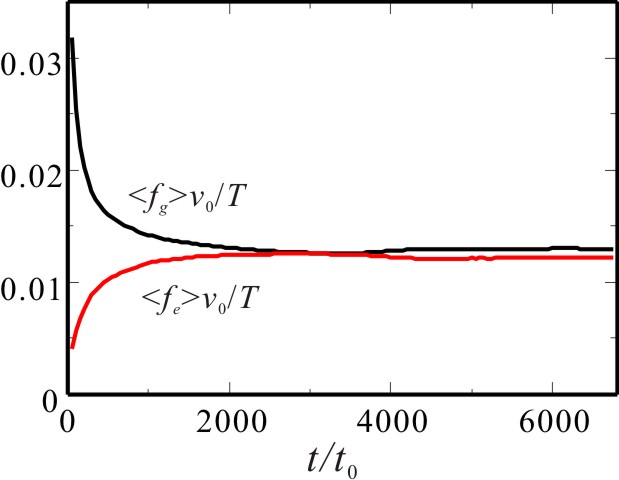}
\caption{
Space averages of  
 $f_gv_0/T $ and 
$f_ev_0/T $ vs time $t$ for $n_ev_0=0.0025$, 
where  $\chi=2.5$ and $\av{\phi}=1/2$.  
}
\end{figure}

\section{Summary and concluding remarks}

In this work, 
we  have presented 
dynamic equations 
for binary mixtures containing ions, where 
the free energy includes the solvation interactions.
(i) As the first application, we have calculated 
the dynamic structure factor $G(q,t)$ in (4.1) 
in one-phase states accounting for the ion motions. 
Its relaxation  is slowed down on approaching 
the spinodal as in (4.12), which occurs at an 
intermediate wave number $q_m$ for $\gamma_{\rm p}>1$.
Here $\gamma_{\rm p}$ is the asymmetry parameter 
of solvation.  It also exhibits a double-exponential 
relaxation in the long wavelength limit 
(in dynamic light scattering) as in (4.19). 
(ii) As the second application, 
we have  numerically 
demonstrated emergence of 
mesophases with addition of 
an antagonistic salt, though 
our simulations are  in two dimensions 
and at the critical composition. 
We have obtained a dramatic 
increase of the structure factor $S(q)$ 
at an intermediate wave number in Figs.2 and 7 in accord with 
the experiment of Sadakane {\it et al.}\cite{Sadakane,S1}.  
In these mesophases, the gradient free energy and 
long-range electrostatic energy are balanced 
as in Figs.5 and 10. 
We have found that 
the surface tension vanishes 
in the mesophase in deep quenching.

The present simulation is still  very preliminary 
and more systematic analysis 
is needed in future work. 
In particular, the phase diagram 
in the parameter space of $\chi$, $\av{\phi}$, 
and $n_e$ is required. 
While  our simulation captures  
some salent features of the neutron 
 scattering experiments \cite{Sadakane,S1}, 
the calculated structure factor 
$S(q)$  cannot be 
compared with the observed 
intensity  quantitatively. 
There are many parameters in our theory 
and we cannot judge whether or not our choice 
in (5.3) is appropriate 
 for the experimental system. 
In particular, 
the solvation parameters $g_1$ and $g_2$ 
are not known for 
mixtures of D$_2$O  and 3MP.

Our simulation suggests  that 
 addition of an antagonistic salt to a binary mixture 
can decrease  the surface tension 
of  a macroscopic liquid-liquid interface even to zero. 
We may then predict  a salt-induced  
interface instability, leading to  
 emulsification. 
We also mention  measurements of the dynamic scattering, 
the electric conductivity, 
 and the rheological properties, 
as new experiments using an antagonistic salt.

\noindent{\bf Acknowledgments}\\
\noindent 
This work was supported by Grant-in-Aid 
for Scientific Research on Priority 
Area ``Soft Matter Physics'' from the Ministry of Education, 
Culture, Sports, Science and Technology of Japan.
Thanks are also due to K. Sadakane and H. Seto for 
informative discussions.

\end{document}